\documentclass{iau-JDSS}
\usepackage{graphicx}

\title[Abundance structure and chemical evolution of the Galactic disc] 
{Abundance structure and chemical evolution of the Galactic disc}

\author[Thomas Bensby]   
{Thomas Bensby$^1$
 \and Sofia Feltzing$^2$}

\affiliation{$^1$European Southern Observatory, Alonso de Cordova 3107, Vitacura, 
Santiago, Chile \\ email: {\tt tbensby@eso.org} \\
$^2$Lund Observatory, Box 43, SE-221\,00 Lund, Sweden \\
email: {\tt sofia@astro.lu.se}}

\pubyear{2009}
\volume{xxx}  
\pagerange{xxx--xxx}
\jname{The Galactic Plane, in depth and across the spectrum}
\editors{Janet Drew, Melvin Hoare, Nicholas Walton, eds.}
\begin{document}

\maketitle

\begin{abstract}
   We have obtained high-resolution, high signal-to-noise spectra
   for 899 F and G dwarf stars in the Solar neighbourhood. The stars
   were selected on the basis of their kinematic properties to trace
   the thin and thick discs, the Hercules stream, and the metal-rich 
   stellar halo. A significant number of stars with kinematic properties
   'in between' the thin and thick discs were also observed in order to
   in greater detail investigate the dichotomy of the Galactic disc.
   All stars have been 
   homogeneously analysed, using the exact same methods, atomic data,
   model atmospheres, etc., and also 
   truly differentially to the Sun. Hence, the sample is likely to be
   free from internal errors, allowing us to, in a multi-dimensional
   space consisting of detailed elemental abundances, stellar ages, and 
   the full three-dimensional space velocities, reveal very 
   small differences between the stellar populations.
\end{abstract}



\begin{figure}[ht]
\centering
\resizebox{\hsize}{!}{\includegraphics[bb=18 144 592 430,clip]{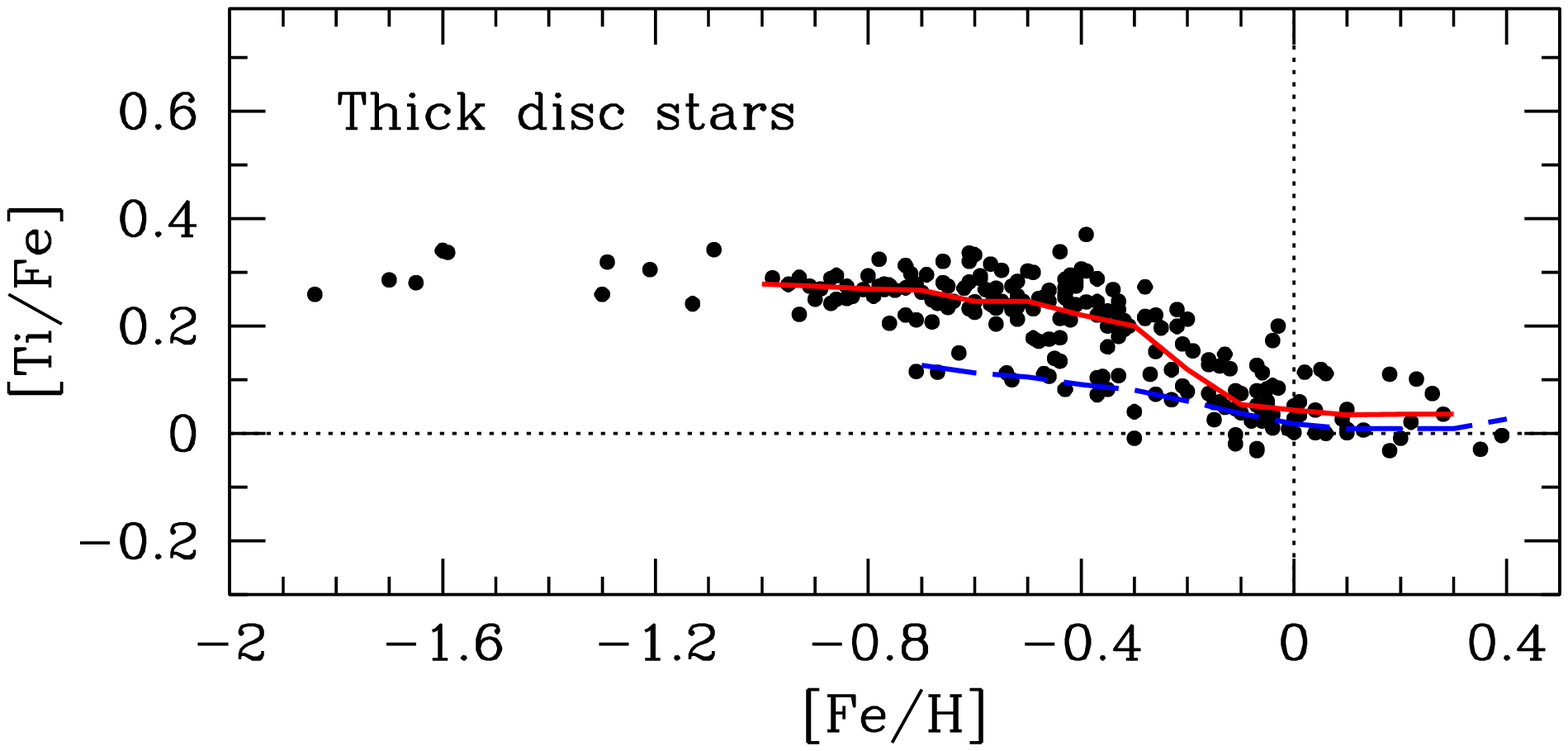}
                      \includegraphics[bb=18 144 592 430,clip]{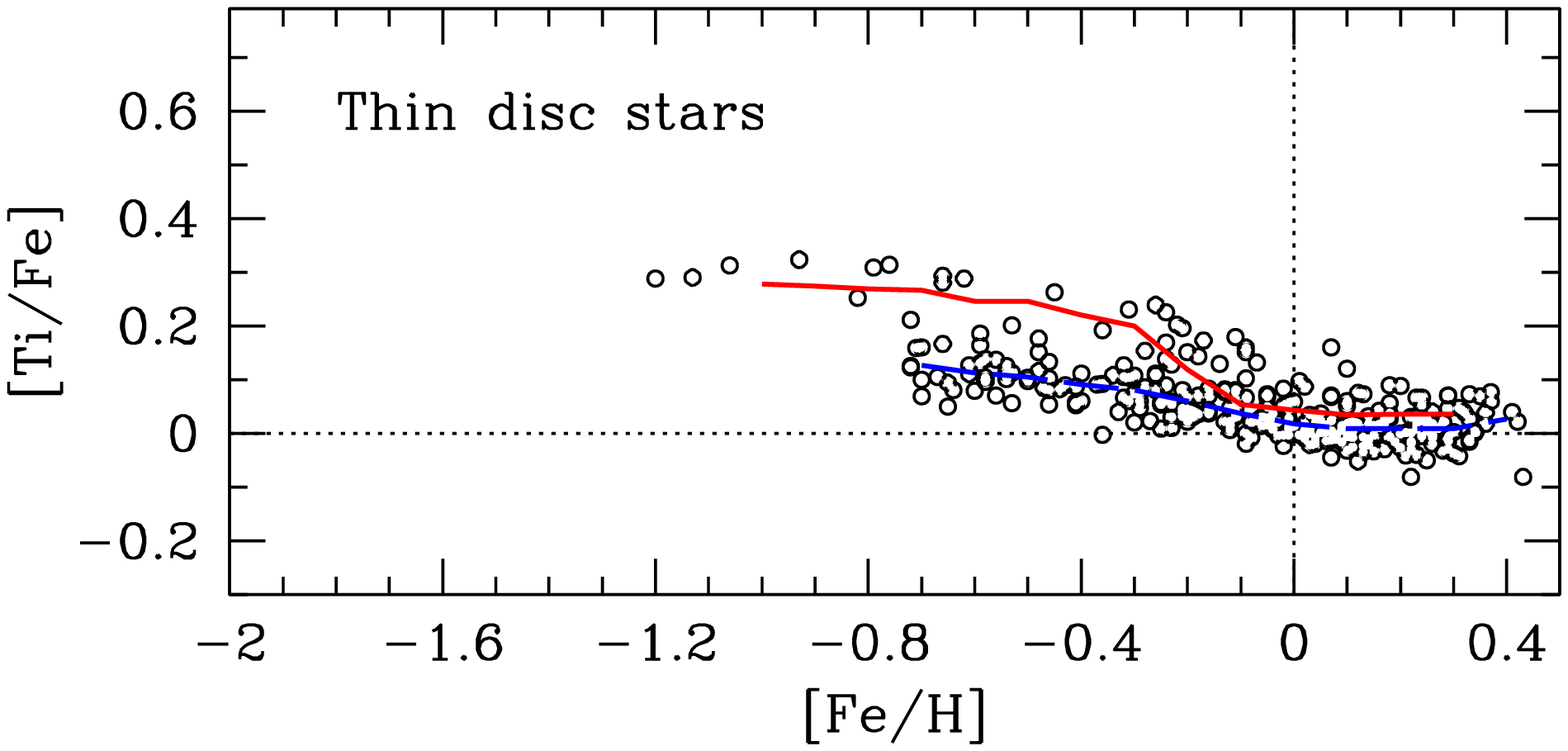}}
\end{figure}
\vspace{-0.5cm}
Compared to our previous studies of the Galactic thin and thick discs
(Bensby et al. 2003, 2005) the current stellar sample is larger by 
a factor of $\sim8$.  The figure above shows
the thin and thick disc abundance trends based on kinematical
selection criteria only. The red full line in each plot is the running
median from the thick disc stars, and the dashed blue line the running 
median from the thin disc stars. It is clear that
there is separation between the two discs up to at least solar
metallicities, signaling the dichotomy of the Galactic stellar disc,
and that the two discs have had very different chemical histories. 
First results, based on this enlarged sample, 
regarding the origin of the Hercules stream and the metal-rich
limit of the thick disc were published in Bensby et al.~(2007a,b). 
The full data set will be published in the fourth quarter of 2009
where we in great detail will investigate the abundance structure
and chemical evolution of the
Galactic stellar disc.



\end{document}